\documentclass[fleqn,12pt]{article}
\usepackage{graphicx}
\usepackage{multicol}
\usepackage{amsmath}
\usepackage{floatflt}

\title{ Frequency Analysis of the noise in the Fowler(n)  
        sampling of a H2RG(2Kx2K) Near-IR Detector  }

\author{G. Smadja$^1$, C. Cerna$^2$, A. Castera$^1$ and A. Ealet$^2$\\
\\
$^1$\normalsize{IPNL, Institut de Physique Nucléaire de Lyon}\\
\small{4, rue Enrico Fermi, 69622, Villeurbanne cedex,France}\\
$^2$\normalsize{CPPM, Centre de Physique des Particules de Marseille}\\
\small{163 avenue de Luminy Case 902, 13288 marseille Cedex ,France}}
 
\begin{document}
\setlength{\textheight}{22cm}
\setlength{\textwidth}{20cm}

\maketitle 

\begin{abstract}
The readout noise of a H2RG HgCdTe NIR detector from Teledyne is 
measured  at a temperature T=110K. It is shown that a Fowler
mode with n = 240 allows to reach a noise of 2.63e (single read).   
A description of the power spectrum in terms of 3 parameters reproduces 
the variation of the noise as a function the number of Fowler samples, 
as well as its dependence on the periodicity of the sampling. The 
variance of the noise decreases with frequency with an effective 
power of 0.62 in our measurement domain. The behaviour of the detector 
under different experimental conditions can then be predicted.

\vspace{1pc}
\end{abstract}

\section{The Apparatus}
The measurements described in this paper  were carried out in a dedicated setup built
to evaluate Hawaii 2RG (HgCdTe) detectors from Teledyne.
The detector was on loan from LBNL in view of the evaluation of its 
performance when used in a spectrograph for the JDEM project \cite{spectro:refspectro}. 
The cryostat can be operated in a range of temperature extending from 110K to 160K 
with fluctuations smaller than 0.1 K, and its equilibrium temperature  
in the absence of heating is about 110K. The polarisation of the substrate was 
chosen as $V_{sub}- V_{reset} = 0.4 $ V in all the data analysed in this work. 
Additional experimental details are provided in a previous paper \cite{smadja:non-linear}
where the conversion factor of our setup (e/ADCU) at T = 110K was evaluated to be 2.042 e/ADU
for a single pixel. 
The goal of the present measurements is to lay the ground for a determination of the
frequency  distribution of the noise, so as to be able to predict the behaviour of
the detector in different experimental conditions (Fowler samplings, interval between 
groups, etc...). The merit of the power spectrum analysis is also the link it provides
with the physical processes in the CMOS semi-conductor, such as the trapping and 
detrapping of electrons with a distribution of time constants. 

\section{The measurement method}

In the 2Kx2K H2RG detector, a window of 31x31 pixels has been selected.
The clock frequency is 100khz, as anticipated in the SNAP-JDEM project,
and the time needed for the readout of 1 frame of 30x30 pixels 
is $\delta = 17.5\hbox{ms}$ . The non destructive frame readouts are organised 
into 'groups' of 250 frames, read at the clock frequency of 100kHz and separated by a
time interval which can be tuned, with the clocking of the pixels stopped. 
Taking into account the readout of the frames, the periodicity of the readout of a 
given pixel between 2 consecutive groups varies from 5s to 13s. The selection of a 
small window allows to reach a higher repetition rate for the frame readout, and 
to save on the overall measurement  time as well as on computing resources. 
All the frames are stored on disk for further analysis. 
The noise is characterised by two variance measurements:\\
-the frame to frame variance, with the readout interval of $\delta = 17.5\hbox{ms}$ for a 'window' frame
 31x31 as considered here, is convenient for a calibration of the detector and 
its readout. It was studied in \cite{smadja:non-linear}. \\
-the Fowler(n) group  to group  variance measured by the spread of the  difference of the averages 
  of 2 consecutive groups for a given pixel will allow to characterize the
  noise performance of the H2RG on longer time scales, and will be investigated here.

The Fowler($n$) noise is evaluated in each pixel from the differences  $D_k$ of the average signal
in groups $k$ and $k -1$.

   
\begin{eqnarray}   
D_k = \frac{1}{n}\sum_{i=1}^{i=n} s_{(t_0+k\Delta+ i\delta)} - s_{(t_0+ (k-1) \Delta + i \delta)}
\end{eqnarray}

The number of frames is 250 in the acquisition, but will be varied from 1 to 240 in the offline
analysis as the first 10 frames of each group  are always ignored. 
The noise is measured by the average of $D_n(\Delta)^2$ over the $N+1$  groups stored during the exposure.
The number of groups in the on-line acquisition was actually 200 in the exposures studied here, but 
the first 10 groups were eliminated from our investigation. 
\begin{flalign}    
\sigma^2  = \frac{1}{N-1}  \sum_{k=1}^{N} \left(D_k - \langle D_k \rangle \right)^2 
\end{flalign}
with
\begin{equation}
\langle D_k \rangle = \frac{1}{N} \sum_{k=1}^{N} D_k
\end{equation}

\begin{floatingfigure}[hp]{9cm}
   \includegraphics[width=9cm]{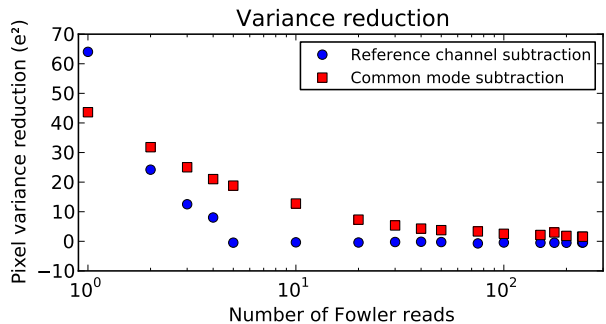}
   \caption{Noise reduction from reference channel and common mode subtractions. The second one is applied on top
of the first one}
\end{floatingfigure}

The  subtraction of the reference channel, which follows synchronously the level of a fixed capacitance, 
and corrects possible bias changes reduces the variance when the number of Fowler samplings $n < 10 $, 
as shown in Figure 1 for a  group periodicity of 5.23s. For larger numbers of reads, the variance of 
the noise is however slightly increased,
by typically $0.1 \hbox{ADU}^2$. We attribute this to low frequency voltage offsets, which are at least
partially cured by the next step. 

\section{Common-mode subtraction at low frequency }

\begin{figure}[htbp]
\begin{center}
\includegraphics[width= 6.5 cm,clip]{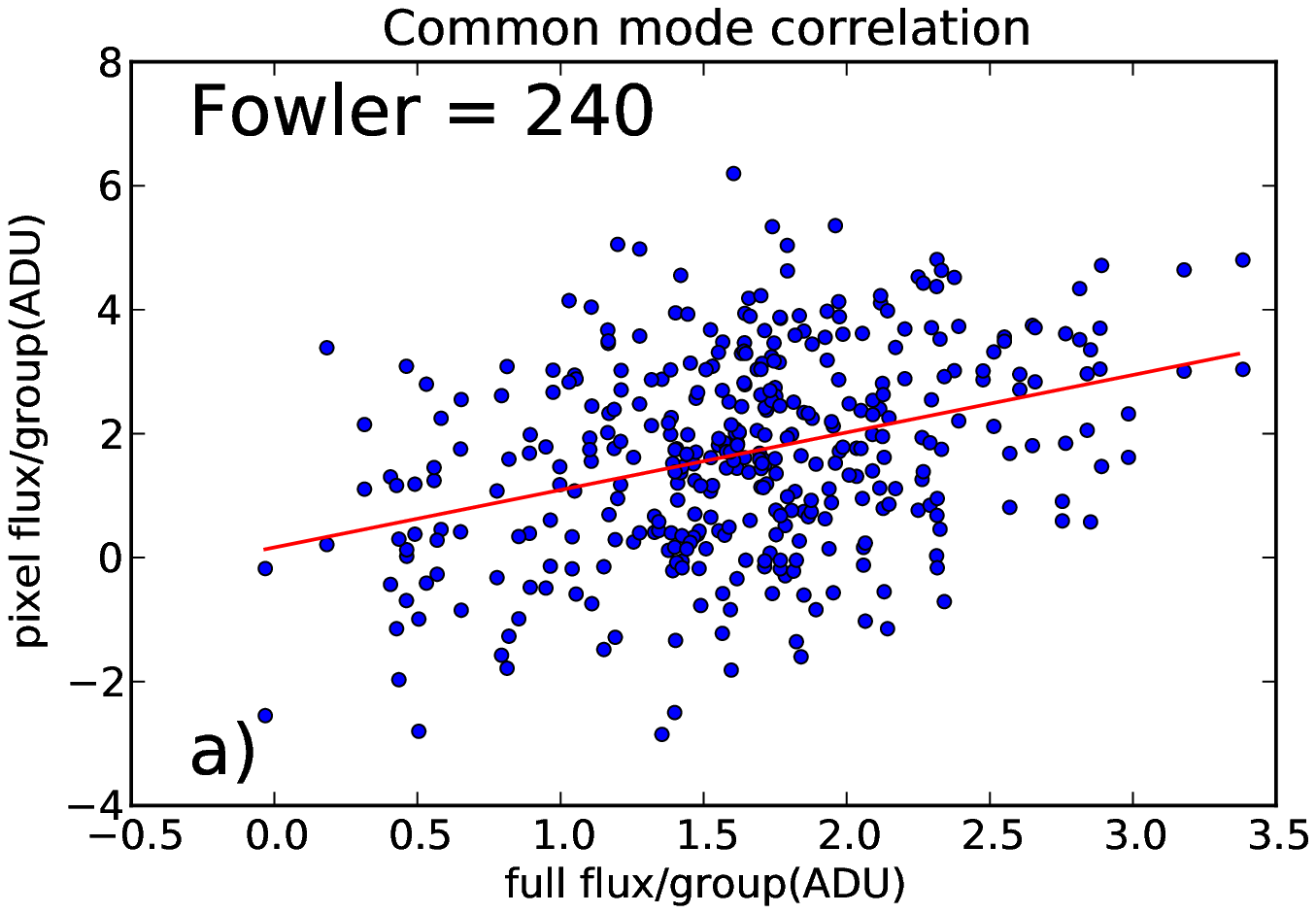}
\hskip .4 true cm
\includegraphics[width= 6.5 cm,clip]{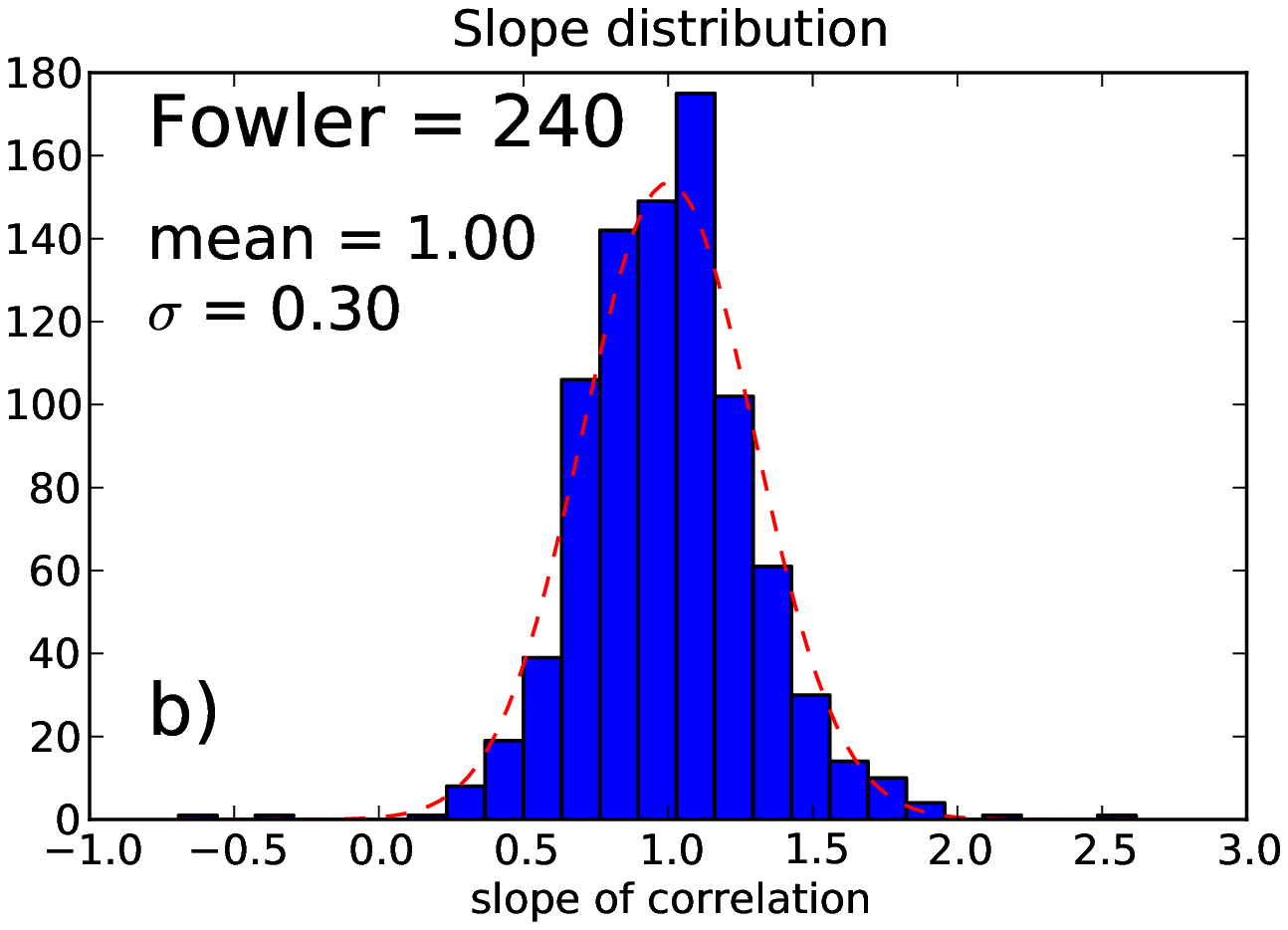}
\caption{a) group to group correlation between pixel flux (Fowler=240)
and full flux: 2 pixels have been selected at random. 
         b) distribution of the correlation slope (Fowler = 240)found for all pixels}
\end{center}
\end{figure}

The reference channel subtraction takes out most of the high frequency common modes, but we still observe 
the impact of common modes remaining at lower frequencies in 
Figure 2a). This figure shows the remaining correlation between the differential flux observed 
for 2 randomly selected pixels between 2 consecutive groups (in the absence of any illumination), 
and the differential flux of the sum over all pixels. 
Both are averaged over the 240 Fowler acquisitions of the group, and the 200 groups of the 
exposure are shown. The value of the slope of the correlation is stored for each pixel, and 
its distribution is shown for all pixels in Figure 2b), where it is seen to be compatible with 
unity within the measurement errors from the noise. 
A 'common' offset for each group is then obtained from a second straight line fit throughout 
the 'full' (integrated over all pixels) observed fluxes of all 200 groups as a function of the group number.
The resulting  variance reduction is shown in Figure 1 as a function of the Fowler number $n$, and the
resulting distribution of the 240 corrected flux differences $D_k$  is shown in fig 3, 
for a typical exposure  at 110K with Fowler samplings n = 240.  The rms of the distribution is 4 electrons.
As this value arises from the difference between the averages over 2 (consecutive) groups of 240 frames, 
it can then be considered that the noise in a single Fowler readout is 2.85 electrons. We shall
see  in section 5 that part of this noise can be assigned to a parasitic photon flux, and that the 
actual performance reached is better. 
\begin{figure}[htbp]
\begin{center}
\includegraphics[width= 6 cm,clip]{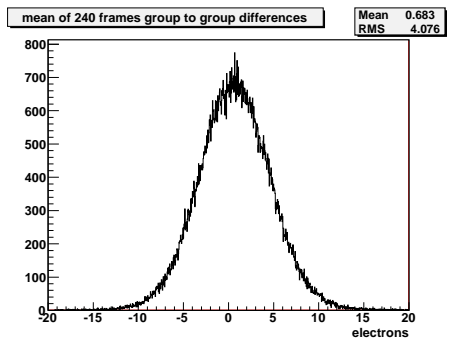}
\caption{group to group difference distribution for Fowler(240)}
\end{center}
\end{figure}
The improvement observed with this common-mode subtraction may be a combination of intrinsic
detector readout properties and of the specific implementation of the outside low voltage supplies.
   
\section{Evolution with the number of samplings }

The evolution of the variance of the difference $D_k$ between consecutive groups as a function of 
the Fowler(n) number is shown in Figure 4 for an intergroup  periodicity of 5.23s after the 
reference channel and common mode corrections have been applied. 
The duration of the full exposure (250 groups of 250 frames each) is 1280 seconds.
The value reached for Fowler(240) corresponds to a noise of 2.85e for a single group read. It is 
apparent that the variance does {\it NOT} decrease like $1/n$, as would be expected for independent 
measurements,which would lead to a group to group noise of 1.6e  at $n = 240$. There are two main 
contributions to these correlations: the non destructive readout itself, and the time domain 
correlations induced by the trapping and detrapping of electrons. 
This behaviour has been noted in several previous publications, such as \cite{Finger:Fowler}, 
\cite{Smith:Fowler}, \cite{Schubnell:Fowler}, and implies the presence
of time dependent correlations, which can be described by a frequency power spectrum. 
The connection between the time constant of trapping sites and the frequency power spectrum
was first stated by \cite{Schottky:trapping}, and a recent presentation can be found in the 
lectures of  \cite{rsasaki:lectures}.

\section{The extra flux}

Before turning to the frequency power spectrum, we must take care of the other contributions to 
the observed variance. Even in the absence of any illumination inside the cryostat, 
The ADC level rises linearly with time, at the rate of 0.6 e/pix/s whatever the group 
periodicity. This rate is not a leakage current, 
as it is (almost) insensitive to the operating temperature, but it could be due either 
to a parasitic photon source or to some baseline drift. We shall assume here that the cause is a photon source
subject to stochastic fluctuations, and we consider that the quantitative agreement with the data 
reached under this assumption  argues in its favour. 
The impact of such a constant photon source on the Fowler(n) variance  is recalled in Appendix A for 
completeness, although it has already been evaluated in the litterature in a slightly different
context, such as in \cite{Rauscher:Poisson}, where the impact on a straight line fit is directly
evaluated. The result is of the form
\begin{equation}
\langle P_n^2 \rangle  = DI + f di_f 
\end{equation}
where $\langle P_n^2 \rangle$ is the flux contribution to the variance, $DI$ the intergroup contribution 
(the flux already included in the $n$ Fowler frames must be omitted), $di_f$
the frame to frame contribution, and the factor $f$ takes into account the correlations arising
from the non destructive readouts, as shown in Appendix A. 
\begin{flalign}
f= \frac{2 di_f}{n^2} \cdot \frac{n(n-1)(2n - 1)}{6}
\end{flalign}   
\section{Frequency spectrum of the noise}

We show  in Appendix B how to relate the Fowler averaging to the autocorrelation products.
As the autocorrelation of the noise is itself directly related to the frequency power spectrum 
$f(\omega)$, the group to group noise for any number of Fowler samplings can then 
be expressed in terms of the power spectrum:
\begin{eqnarray} 
\langle s_{t}  s_{t+\tau} \rangle = \int_{\omega 1} ^{\omega 2} d\omega cos(\omega \tau) f(\omega) 
\end{eqnarray}
The power spectrum $f(\omega)$ will be assumed to be of the form
\begin{equation}
f(\omega) = A + \frac{B}{\omega^{\alpha}}
\end{equation}
If this functional dependance were exact, the white 'thermal' amplitude $A$ would be a positive number, 
and $\alpha$ might generaly be expected to be close to 1. Our measurements are however performed 
in a frequency range where the white noise is subdominant, so that A is just a parameter of 
the fit, and it turns out it will be found to be small and negative. The power $\alpha$ found is also 
an effective power in our frequency range, which may combine the detector and FET effects.  
The merit of this 3 parameter formalism is that it describes 
the shape of the Fowler (n) function with an accuracy of ~2\% and has a predictive 
power for different group or frame periodicities as well.  

As shown in Appendix  B, the measured variance $\langle D_n(\Delta)^2 \rangle$ can be expressed as a sum of 
autocorrelation products, and an explicit expression of the integral relating 
the power spectrum $f(\omega)$ to the measured variance
$\langle D_n(\Delta)^2 \rangle$ is  derived in the appendix:
\begin{eqnarray}
\langle D_n^2 \rangle &  = &   \int d\omega(1-cos\omega \Delta)(A + \frac{B}{\omega}) \nonumber \\
        & \cdot & \left[ \frac{2}{n} + \frac{\mathcal{M}}{n sin(\omega \delta/2)} \right. + \left. \frac{\mathcal{N}}{n^2 sin^2(\omega \delta/2)} \right] 
\end{eqnarray}  
With :
\begin{equation*}
\mathcal{M}=4cos\left(\frac{n \omega \delta}{2}\right) sin\left(\frac{(n-1) \omega \delta}{2}\right)
\end{equation*}  
and
\begin{equation*}
\mathcal{N}=1 - cos(n \omega \delta)-2n sin\left(\frac{\omega(2n-1)\delta}{2}\right)sin\left(\frac{\omega \delta}{2}\right)
\end{equation*}  

In this relation, $\delta$ and $\Delta$ are the frame to frame and group to group periodicities. 
The predicted variance for given values of $A$,$B$, and the power $\alpha$,
are evaluated at 10 different Fowler samplings (to spare computer time) and 
they are adjusted to the data for 2 different group periodicities   of 5.23s
and 13.11s by minimizing the $\chi^2$.  
Given the very high statistical accuracy of the measurements, and their reproducibility, the 
typical computed errors on each data point in Figure 4 is 3 to 4 $10^{-3}$. 
On the other hand, the 3 parameter model used for the power spectrum reproduces 
the data to an accuracy of 1 to 2\% for the best fits. The errors were rescaled 
to match this difference and follow the observed trend, as seen in Table 1.    

\begin{table}[!htb]
\caption{ \it Variance evolution with the number of Fowler samplings)}
\label{table:1}
\newcommand{\m}{\hphantom{$-$}}
\newcommand{\cc}[1]{\multicolumn{1}{c}{#1}}
\renewcommand{\tabcolsep}{0.99pc} 
\renewcommand{\arraystretch}{1.2} 
\begin{tabular*}{1.01 \textwidth}{c c c c c }
\hline
\multicolumn{5}{c}{Intergroup periodicity 5.23s, exposure time = 1280s }\\
\hline
Fowler nb.   & Adjusted  noise & Meas. noise  &   Meas.  error & Assigned error\\
   (n)          & (e)             & (e)            &   (e)             &  (e)    \\
\hline
2           & 18.58      & 18.28      & 0.135     & 0.37\\
4           & 14.37      & 14.61      & 0.055     & 0.29\\
5           & 13.29      & 13.58      & 0.033     & 0.27\\
10          & 10.57      & 10.81      & 0.022     & 0.22\\
20          &  8.55     & 8.62       & 0.025     & 0.086\\  
50          &  6.56     & 6.51       & 0.045     & 0.065\\
100         &  5.37     & 5.34       & 0.050     & 0.053\\
150         &  4.75     & 4.74       & 0.040     & 0.047\\
175         &  4.52     & 4.52       & 0.030     & 0.045\\                 
200         &  4.32     & 4.34       & 0.023     & 0.043\\
240         &  4.04     & 4.08       & 0.017     & 0.041\\  
\hline
\multicolumn{5}{c}{intergroup periodicity 13.11s, exposure time = 3275s}\\
\hline
Fowler nb.   & Adjusted  noise & Meas. noise  &   Meas. error & Assigned error\\
   (n)          & (e)             & (e)            &   (e)             &  (e)    \\
\hline
2           &  19.04     & 18.72      & 0.025     & 0.37\\
4           &  15.02     & 15.26      & 0.032     & 0.30\\
5           &  13.98     & 14.27      & 0.050     & 0.28\\
10          &  11.35     & 11.56      & 0.064     & 0.23\\
20          &  9.40    & 9.40       & 0.040     & 0.094\\  
50          &  7.43     & 7.35       & 0.037     & 0.073\\
100         &  6.30     & 6.26       & 0.032     & 0.063\\
150         &  5.74     & 5.73       & 0.031     & 0.057\\
175         &  5.54     & 5.55       & 0.030     & 0.055\\                 
200         &  5.37     & 5.40       & 0.032     & 0.054\\
240         &  5.15     & 5.20       & 0.031     & 0.052\\  
\hline

\end{tabular*}    
\end{table}
The best fits differ for the 2 data samples, although  the power spectra
found are close to each other.  
The estimated increase in the noise at Fowler=240 is shared equally 
(in terms of variance) between the stochastic contribution  and the 1/f contributions.

\begin{table}[!htb]
\caption{\it Adjusted power spectrum parameters at 5.23 and 13.11s group periodicities}
\label{table:2}
\newcommand{\m}{\hphantom{$-$}}
\newcommand{\cc}[1]{\multicolumn{1}{c}{#1}}
\renewcommand{\arraystretch}{1.2} 
\begin{tabular*}{1.01 \textwidth}{c c c c c }
Exposure time/   & Thermal  & 1/f    &   1/f power  & $\chi^2$/Ndf\\
   period(s/s)          &         A ($\mu$V$^2$Hz$^{-1}$)    &  B ($\mu$V$^2$Hz$^{-1}$) & $\alpha$  &     \\
\hline

1280/5.23         &    8.445    & 734.819     & 0.62     & 6.063/8 \\
3280/13.11        &   -1.525   &  846.563     & 0.57     & 6.000/8 \\ 

\hline
\end{tabular*}    
\end{table}

The (dominant) B contribution amounts to 27.107e/(Hz)$^{-1/2}$ (5s intergroup) and 29.096e/(Hz)$^{-1/2}$ (13s intergroup).
The result of the fit is seen in Figure 4a, and the estimated errors on $A$ and $B$ shown
as contour plots in Fig 4b) are only indicative, as the errors are just set so that the $ \chi^2/Ndf$  would be of 
the order of unity. 

\begin{figure}[htbp]
\begin{center}
\includegraphics[width= 6.5 cm,clip]{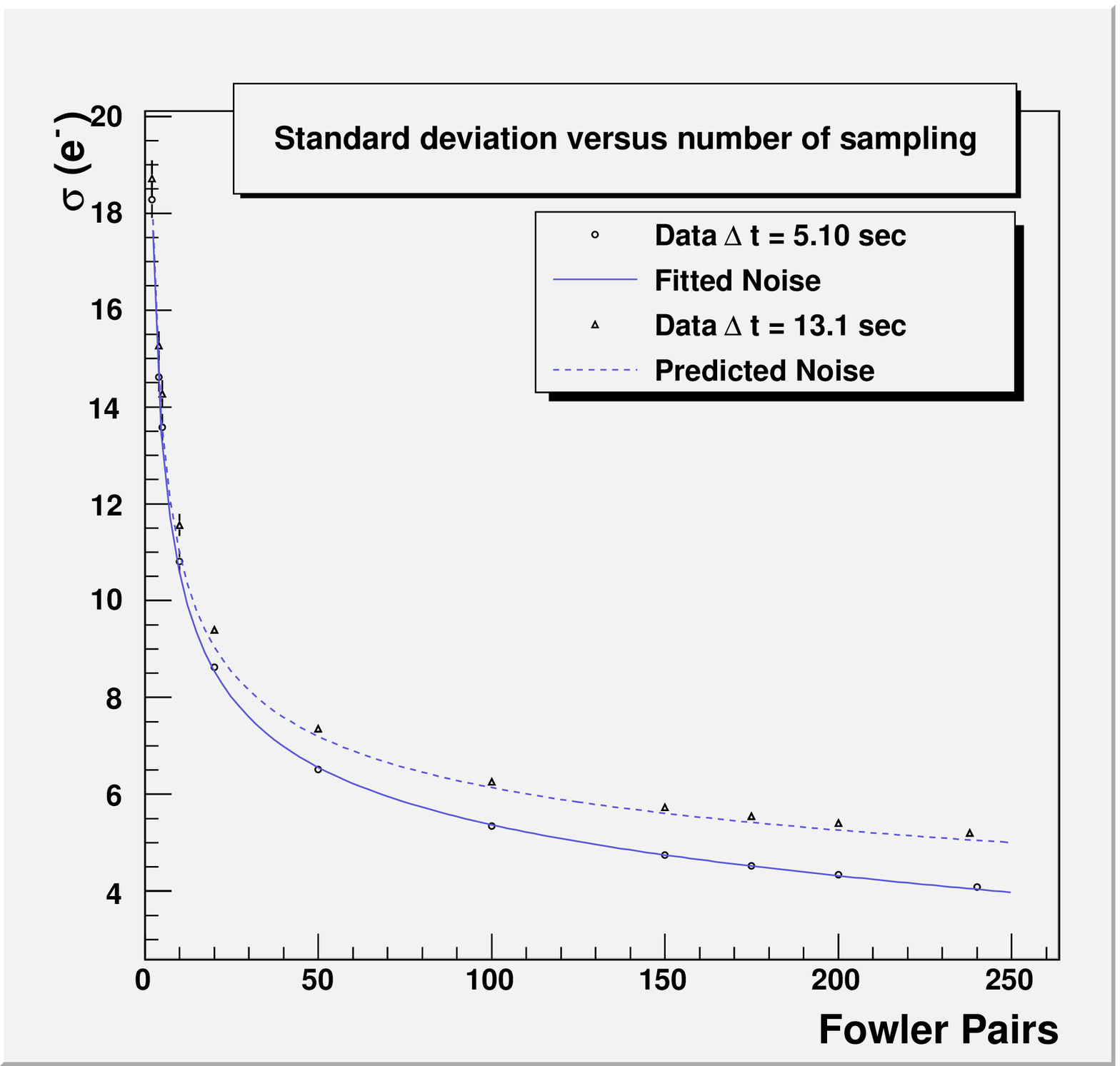}
\hskip 0.5 true cm
\includegraphics[width= 6.5 cm,clip]{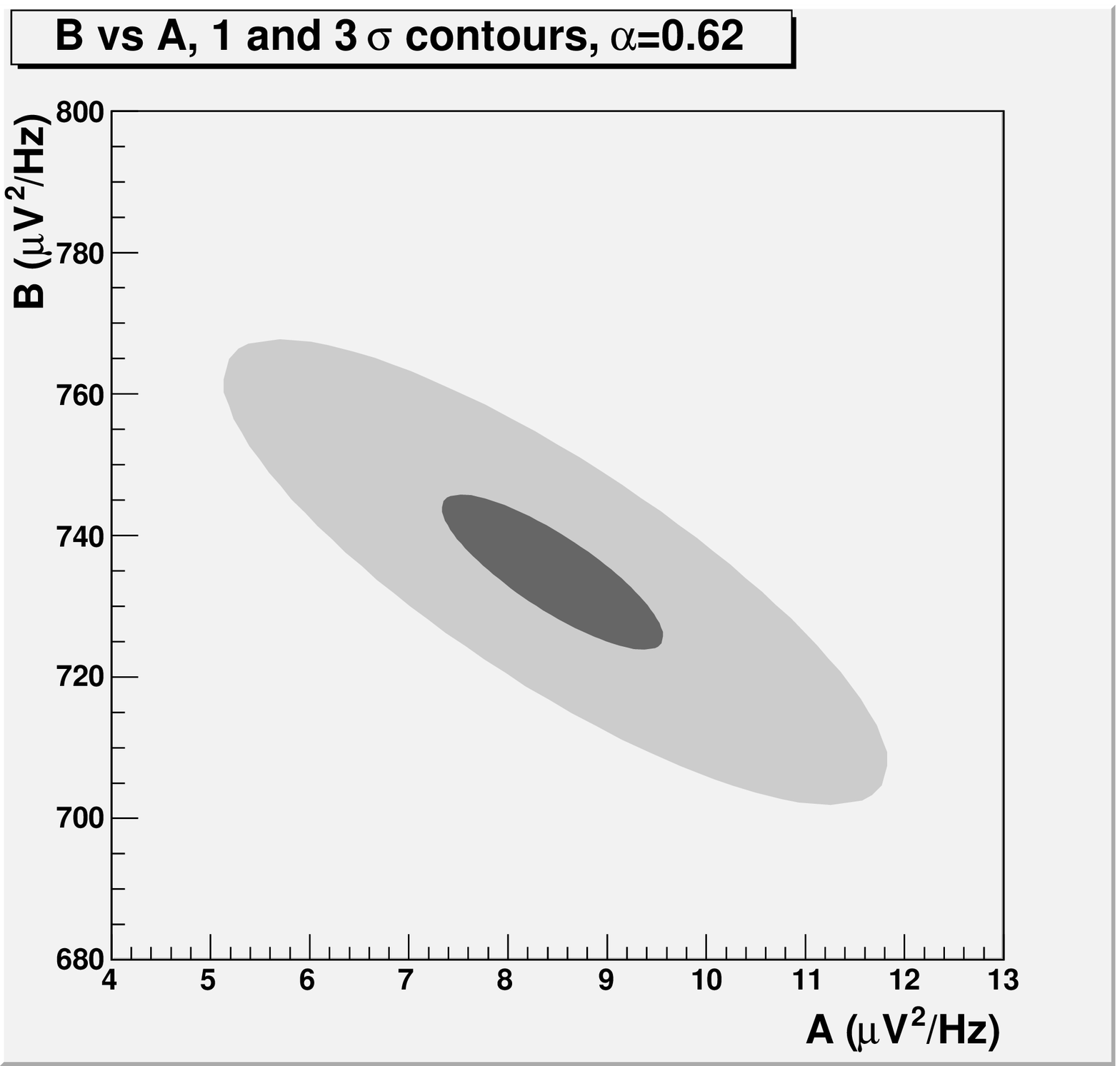}
\caption{a) Fowler (n) noise measurement and adjusted frequency spectrum  at a group
periodicity of 5.23s 
         b) contour plot of the A and B errors }
\end{center}
\end{figure}

\begin{figure}[htbp]
\begin{center}
\includegraphics[width= 9. cm,clip]{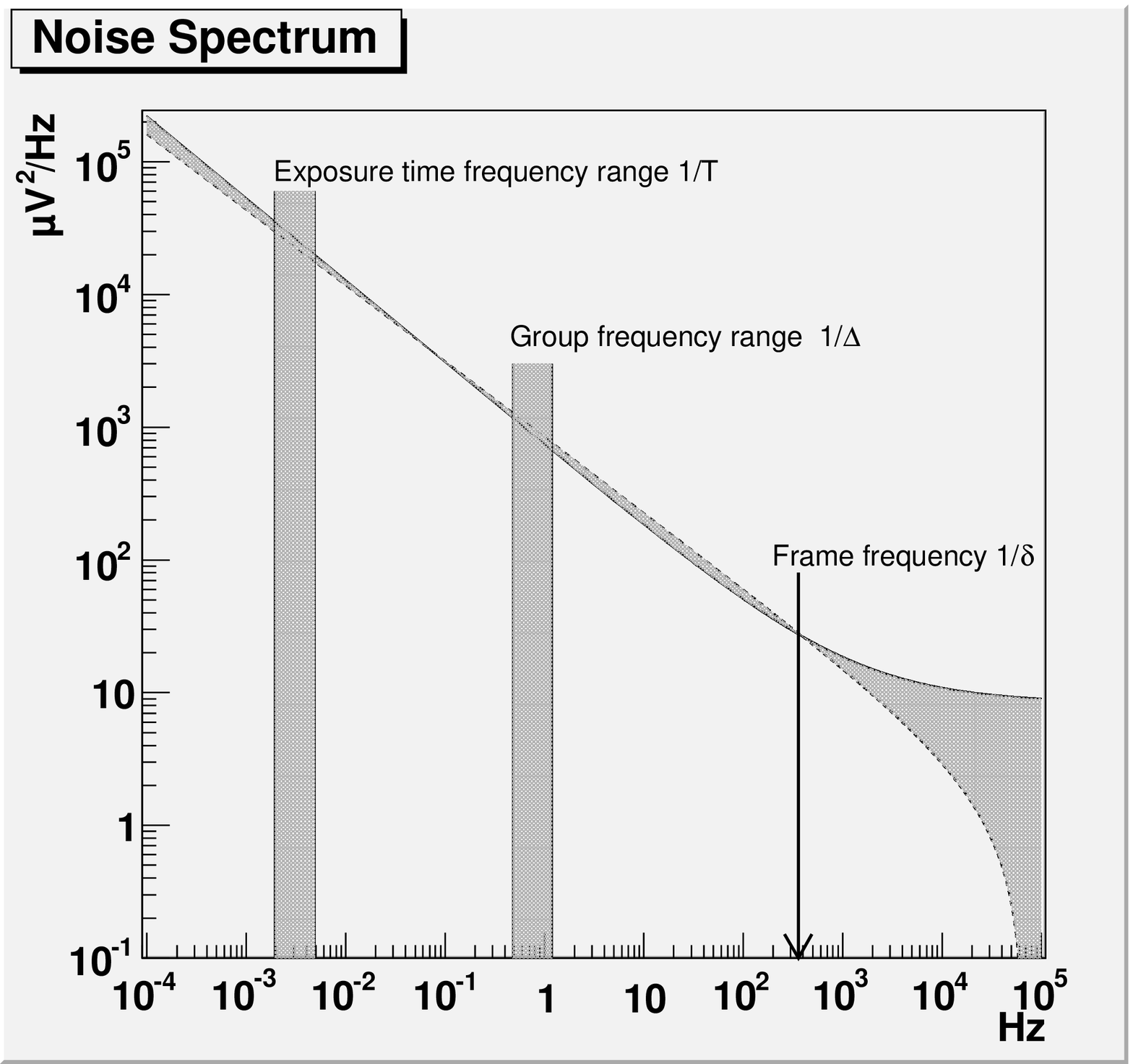}
\caption{Frequency power spectrum of the noise with the adjusted parameters at
group periodicities of 5.23 (continuous line) and 13.11s (dashed)} 
\end{center}
\end{figure}

\section{Prediction of the group periodicity dependence}

We can also check how accurately the evolution with the group period is 
predicted from the restricted lever arm in frequency provided by the 
Fowler(n) dependence of the noise at a single delay. 
We show in table  3 the dependence of the variance expected with 240 Fowler samples 
on the intergroup delay. The stochastic Poisson contribution from the assumed 
'extra flux'  is  substantial, and the agreement with the observations suggests that 
the general features of the physical contributions 
to the noise are understood. A small discrepancy is observed at a group periodicity 
of 13s, confirmed by an inspection of the full Fowler prediction in Figure 4a). We are 
tempted to assign this effect to a frequency dependence
of the power $\alpha$, which will be investigated in a later work. 

\begin{table}[!htb]
\caption{\it Variance evolution with group periodicity for Fowler(240)}
\label{table:3}
\newcommand{\m}{\hphantom{$-$}}
\renewcommand{\arraystretch}{1.2} 
\begin{tabular*}{1.01 \textwidth}{c c c c c}
Period/Exp. time & Var. (e$^2$) &   Var.  (e$^2$)&     Var.  (e$^2$)&  Var. (e$^2$)\\
   (s)/(s)       & (flux)   &  (frequency) & (predicted)    &  measured \\
\hline
5.23/1280 & 2.396   &  13.764    & 16.160   & 16.646 \\
6.578/1315 & 3.427   &  14.949   & 18.376 & 18.768 \\
7.304/1461 & 3.941   &  15.452   & 19.393  & 19.564 \\
8.756/1751 & 4.905   &  16.085   & 20.991  & 21.328 \\
10.208/2042 & 5.956   &  16.667    & 22.623   & 23.088 \\  
13.112/3275 & 8.066   &  17.472   & 25.538 & 27.041 \\                       

\hline
\end{tabular*}    
\end{table}
It is seen that the actual readout contribution to the group to group variance is 13.8 $e^2$,  
i.e. 3.71 electron for the difference between 2 Fowler(240) readouts and a group periodicity 
of 5.23s .    
The extrapolation of the parameters adjusted at an intergroup period of 5.23s is compared
to the full Fowler(n) curve at a period of 13.11s in Figure 4a. It is seen that the predicted
noise slightly undersestimates the measured one, as also observed in table 3. The adjustment at
a period of 13.11s has an unphysical  negative value of the white noise component, as seen in Table 2
and Figure 5a). This feature has a minor impact as all the experimental data is carried in
a frequency range smaller than  370 Hz, where the 2 frequency spectra are close to each
other.   

\section{Conclusion}
We have shown that the combined use of the reference channel and a common-mode correction lead
to a significant reduction of the detector noise. The second correction can be performed with
the help of non illuminated pixels (unavailable in the window mode) in the  standard operation
of the H2RG detector. With the H2RG lot \# 40 which we measured, we reach a noise of 
2.63e (single read) with Fowler(240) in a 1250s exposure. This is degraded to a noise of
 2.96 e for a 3280s  exposure with an intergroup periodicity of 13s.   
A description of the power spectrum of the noise in terms of 3 parameters also accounts 
quantitatively for the Fowler(n) variation of the noise at different group periodicities
within an accuracy of 2\%. The power law exponent found to reproduce the data is about 0.62.
This 'effective' power law is only approximate, and the impact of additional parameters 
improving the description  of the power spectrum will be investigated in a later 
work, as well as the effect of different polarisation voltages.  

\section{Acknowledgements}
We thank all the institutions who have supported us during this work: Universit\'e Claude Bernard Lyon 1,
The IN2P3/CNRS institute, and the engineers and technicians at IPNL and CPPM who have contributed to
the apparatus F. Charlieux, J. C. Ianigro, P. Karst, and in particular C. Girerd who has designed 
the readout electronics.
We are indebted to C. Bebek (LBNL) for lending us the H2RG detector, and to G. Tarle, M. Schubnell, and 
R. Smith for many questions and suggestions.

\newpage
\begin{center}
{\Large \bf Appendix}
\end{center}
\appendix
\section{Flux and  leakage current contribution}

$\delta$ and $\Delta$ will be the interval betweeen frames and between groups. 
In the Fowler mode, an average is taken between signals
at $t ,t+\delta,t+2 \delta,... t+(n-1) \delta$
and the difference is formed with the homologous signal values at
$t + \Delta,t+\Delta+\delta,t+\Delta+ 2 \delta,... t \Delta +(n-1) \delta$.
The contribution of the stochastic fluctuation is evaluated next
 $DI$ and $di$ will be the mean intergroup and frame to frame fluxes, 
$\delta (DI)$ will stand for the intergroup signal fluctuation, and $\delta_i(k)$ will 
be the flux fluctuation of group $i$ at frame $k$ (the fluctuation of the difference
between frames $k$ and $k-1$). 

\begin{eqnarray*}
S_1 &=& DI + \delta(DI)\\ 
S_2 &=& DI + \delta(DI)+ di +\delta_2(2) - \delta_1(2)\\
S_3 &=& DI +  \delta(DI) + 2di + \delta_2(2)+ \delta_2(3)   -\delta_1(2) - \delta_1(3)\\
S_4 &=& DI +  \delta(DI) + 3di + \delta_2(2) + \delta_2(3) +\delta_2(4)  
           -  \delta_1(2) - \delta_1(3) -\delta_1(4)\\
S_n &=& DI +  \delta(DI) + (n-1)di + \delta_2(2) + ...\delta_2(n) 
                                     - \delta_1(2) -...\delta_1(n) \\ 
\end{eqnarray*}
The average is formed

$$S = \frac{S_1 + S_2 + S_3 + ... S_n}{n}$$

The fluctuating Poisson contribution to the result is 
$$ \frac{1}{n} (\delta_2(2)-\delta_1(2))*(n-1) +  (\delta_2(3)-\delta_1(3))*(n-2)
   + (\delta_2(4)-\delta_1(4))*(n-3) +... +(\delta_2(n)-\delta_1(n))*1$$
    
The variance of $\delta_2(i)$ and   $\delta_1(i)$ are all equal
to the number of electrons between 2 frames $di_f$ (in number of electrons)

$$\delta(S)^2 =  DI + \frac{1}{n^2} 2 di_f^2 (1 + 2^2 +3^2 +.... (n-1)^2)$$

where the sum of the $n-1$ first squares appears\\
$$ 1^2 + 2^2 + 3^2 +...(n-1)^2 = \frac{(n-1)(n)(2n - 1)}{6}$$

The stochastic contribution to the variance Fowler(n) variance  will be 
(in electron units)
\begin{equation}
\delta(S)^2 =  DI + \frac{1}{n^2} 2 di_f^2 \frac{(n-1)(n)(2n - 1)}{6}
\end{equation}
To be added quadratically to the frequency spectrum noise. It is seen that the stochastic error
on the group average grows like the number of Fowler averages, and will exceed the readout error at 
some point. High Fowler numbers harm at large fluxes.

\section {The Wiener relation}
Time domain correlations of the noise are induced by the interplay of the  physical
properties of semi-conductors.  As shown by Wiener, they can be described by the frequency 
power spectrum of the noise,which happens to have a quasi universal behaviour as a
function of frequency, of the form $A + B/\omega^{\alpha} $.  
Assume
$$ x(t) = \int d\omega f(\omega)e{-i \omega t} $$

\begin{eqnarray*}
<x(t) x(t+\Delta)> & = &\lim_{T \rightarrow \infty} \frac{1}{2T}\int_{-T}^{+T} dt d\omega_1 d\omega_2 e^{-i \omega_1 t} e^{i\omega_2 t + \Delta} f(\omega_1) f^{\ast}(\omega_2)\\
                   &   = &\lim_{T \rightarrow \infty} \frac{2 \pi}{2T} \int d\omega e^{i\omega \Delta} f(\omega) f^{\ast}(\omega)
\end{eqnarray*}

As the result should be real
$$f(-\omega) = f(\omega)^{\ast}$$

$$<x(t) x(t+\Delta)> = \frac{\pi}{T}(\int_0^\infty d\omega f(\omega) f(\omega)^{\ast} e^{i\omega \Delta} + 
                                      \int_{-\infty}^0 d\omega f(\omega) f(\omega)^{\ast}e^{i\omega \Delta})$$

So that taking advantage of the reality relation
\begin{eqnarray*}
<x(t) x(t+\Delta)> & = &\frac{\pi}{T} \int_0^\infty d\omega ( e^{i\omega \Delta}+ e^{-i\omega \Delta})
                       f(\omega) f(\omega)^{\ast} \\
                     & = &\frac{2 \pi}{T} \int_0^\infty d\omega cos (\omega \Delta )
                       f(\omega) f(\omega)^{\ast} 
\end{eqnarray*}

including the factor  $2 \pi/T $ in the normalisation of the power spectrum $f$
\begin{equation}
<x(t) x(t+\Delta)> = \int_0^\infty d\omega (\omega)cos (\omega \Delta)
                       f(\omega) f(\omega)^{\ast} 
\end{equation}
The shape of the power spectrum $\mid f(\omega) \mid^2$  will be assumed to be of the form 
$$ \mid f(\omega) \mid^2 = A + \frac{B}{\omega^{\alpha}} $$
where $A$ is the intensity of the white noise component  and $B$ related to the strength of
the $1/\omega$ (1/f) contributions. 

\section{The group to group averages}

The measured quantity is the variance group to group differences in the
Fowler(n) averages. We introduce the time intervals   $\delta$  between 2 reads of 
the same pixel in consecutive frames, and the equivalent interval $\Delta$ 
between two readouts of the same pixel in the same frame in consecutive 
groups.

For the consecutive groups $k$ and $k+1$
\begin{equation}
D_k = \frac{1}{n}(x(t_k+\Delta) -x(t_k) + x(t_k+\delta + \Delta) -x(t_k + \delta) + 
                    x(t_k + 2 \delta + \Delta) -x(t_k + 2 \delta) + etc...) 
\end{equation}
\begin{equation}
<D_k> = \frac{1}{N} \sum_k D_k  
\end{equation}
is the time average of $D$ (i.e. the fluence). The noise variance will be defined by  
\begin{equation}
<D_k^2> = \frac{1}{N} \sum_k (D_k - <D_k>)^2$$
\end{equation}
\begin{eqnarray*}                         
\lefteqn{<D_k^2>=}\\
& & \frac{1}{n^2} (2n <x^2> - 2 n <x(t) x(t+ \Delta)> \\ 
& & +2<x(t+\Delta) x(t + \Delta + \delta)> + 2<x (t+\Delta) x(t + \Delta + 2 \delta)> +...\\
& & -2<x(t+\Delta)x(t+\delta)> -2<x(t+\Delta) x(t+2 \delta) -2<x(t+\Delta) x(t+3\delta)>-...\\
& & -2<x(t)x(t+\Delta+\delta)> -2<x(t)x(t+\Delta+2\delta) < -2 <x(t)x(t+\Delta + 3\delta)>-...\\
& & +2<x(t)x(t+\delta)> + 2<x(t)x(t+2\delta)> + 2<x(t)x(t+3\delta)>+...) \\
& & (n-1) \;similar \;terms \\  
& & +2<x(t+\Delta+\delta) x(t + \Delta + 2\delta)> + 2<x (t+\Delta+ \delta) x(t + \Delta + 3 \delta)> +...\\
& & -2<x(t+\Delta + \delta)x(t+2\delta)>  -2<x(t+\Delta+\delta) x(t+3 \delta)\\ 
& & \qquad \qquad \qquad \qquad \qquad \qquad \;\;\;\;\;-2<x(t+\Delta+\delta) x(t+4\delta)>-...\\
& &  -2<x(t+\delta)x(t+\Delta+2\delta)>  -2<x(t+\delta)x(t+\Delta+3\delta)> \\
& & \qquad \qquad \qquad  \qquad \qquad \qquad \;\;\;\; \;-2 <x(t+\delta)x(t+\Delta + 4\delta)>-...\\
& & +2<x(t+\delta)x(t+2\delta)> + 2<x(t+\delta)x(t+3\delta)> + 2<x(t+\delta)x(t+4\delta)>+...)\\
& &(n-2)\;similar \;terms \\  
& &(...) The\;summation \;is \;continued \;in \;the \;same \;way \\       
\end{eqnarray*}
The correlated products can be explicitly expressed in terms of the power spectrum as in the previous section\\
\begin{eqnarray*}
<x^2> & = & \int d\omega (A + \frac{B}{\omega^{\alpha}})\\
<x(t) x(t+ \Delta)> & = & \int cos(\omega \Delta) (A + \frac{B}{\omega^{\alpha}})\\
\end{eqnarray*} 
And using the time translation invariance\\
\begin{eqnarray*}
<x(t+\Delta) x(t + \Delta + \delta)> & = &  \int cos(\omega \delta) (A + \frac{B}{\omega^{\alpha}}) \\
<x(t+\Delta)x(t+\delta)> & = &  \int cos(\omega (\Delta - \delta)) (A + \frac{B}{\omega^{\alpha}})
\end{eqnarray*} 
etc....
Bringing together the correlated products with the same value
\begin{eqnarray*}
<D_k^2> & = &\frac{2n}{n^2}\int d\omega (1 -cos(\omega \Delta) (A + \frac{B}{\omega^{\alpha}}) \\
        & + & (n-1)(2 cos(\omega \delta) -cos \omega (\Delta-\delta) -cos \omega(\Delta + \delta))\\ 
        & + & (n-2)(2 cos \omega 2 \delta-cos\omega(\Delta - 2 \delta) - cos\omega(\Delta + 2\delta))\\
        & + & (n-3)(2 cos \omega 3 \delta -cos\omega(\Delta - 3 \delta) - cos\omega(\Delta + 3\delta)\\    
  &    & (...) \;the \;summation \;is \;continued \;in \;the \; same\;way \\
\end{eqnarray*}
So that 

\begin{eqnarray}
\lefteqn{<D_k^2> = } \nonumber \\
& & \frac{1}{n^2} [2n \int d\omega(1-cos\omega \Delta)(A + \frac{B}{\omega^{\alpha}})   \nonumber\\ 
        &    & +2(n-1) (\int d\omega 2 cos(\omega \delta) (1 - cos \omega \Delta)( A + \frac{B}{\omega^{\alpha}})
\nonumber\\
        &    & +2(n-2) (\int d\omega 2 cos(2\omega \delta )(1 - cos \omega \Delta)( A + \frac{B}{\omega^{\alpha}})
\nonumber\\ 
        &    & +2(n-3) (\int d\omega 2 cos(3\omega \delta )(1 - cos \omega \Delta)( A + \frac{B}{\omega^{\alpha}})
\nonumber\\
        &    & +......]
\nonumber\\
\end{eqnarray} 

It is straightforward to provide an analytical expression of the sums  in equation (14). 
\section{The summed results}
The summation in equation (14) involves to different series.
\begin{eqnarray}
\lefteqn{cos(\omega \delta) + cos (2 \omega \delta) + ... + cos((n-1) \omega \delta) }  \nonumber\\
& & = Re \{ e^{i \omega \delta} (1 + e^{i\omega \delta } + e^{2 i \omega \delta} +... e^{i(n-2)\omega \delta}) \} 
\nonumber \\
& & = Re \{ e^{i \omega \delta} \frac{1 - e^{i(n-1)\omega \delta}}{1 -e^{i\omega \delta}} \}
\end{eqnarray}

Similarly

\begin{eqnarray}
\lefteqn{cos(\omega \delta) + 2cos (2 \omega \delta) + ... + (n-1) cos((n-1) \omega \delta)} \nonumber \\
& & = \frac{1}{\delta}\frac{\partial}{\partial \omega} (sin(\omega \delta) + sin(2 \omega \delta) + ... + sin(n-1) \omega \delta)  \nonumber \\
& & = \frac{1}{\delta}\frac{\partial}{\partial \omega} Im(e^{i \omega \delta} (1 + e^{i\omega \delta } + e^{2 i \omega \delta} +... e^{i(n-2)\omega \delta})  \nonumber \\
& & = \frac{1}{\delta}\frac{\partial}{\partial \omega} Im\{e^{i \omega \delta} \frac{(1 - e^{i(n-1)\omega \delta})}{1 -e^{i\omega \delta}} \} \nonumber \\
\end{eqnarray}

The Fowler(n) averages expressions can then be derived

\section{Explicit result for the real part}

$$(1 - e^{i\omega \delta}) = 1 - cos \omega \delta -i sin \omega \delta = 
2 \sin^2 (\omega \delta/2) -2i sin \omega \delta/2 cos \omega \delta/2)$$
$$  (1 - e^{i\omega \delta}) = -2i sin(\omega \delta/2) e^{i\omega \delta/2}$$

$$ \frac{e^{i \omega \delta}}{1-e^{i \omega \delta}} = \frac{e^{i \omega \delta}}{-2isin (\omega \delta/2) e^{i\omega \delta/2}} =  \frac{i}{2} \frac{e^{i \omega \delta/2}}{sin (\omega  \delta/2)}    $$

The cosine summation is 

$$
\frac{4}{n}Re(\frac{i e^{i\omega \delta/2}}{sin(\omega \delta/2)}(1 - e^{i \omega(n-1)\delta/2)})
$$
As
$$(1 - e^{i\omega (n-1)\delta}) = -2i sin(\omega (n-1) \delta/2) e^{i\omega (n-1)\delta/2}$$
 
The real part contribution to the variance $<D_k^2>$ is 
\begin{equation}
\frac{4}{n}  \frac{cos(n \omega \delta/2) sin((n-1) \omega \delta/2)}{sin(\omega \delta/2 )} 
    (1-cos \omega \Delta) (A + \frac{B}{\omega^{\alpha}})
\end{equation}
\section{Explicit results for the imaginary part}

$$-\frac{4}{n^2}
 \frac{1}{\delta}\frac{\partial}{\partial \omega} Im(e^{i \omega \delta} \frac{(1 - e^{i(n-1)\omega \delta)}}{1 -e^{i\omega \delta}} )$$
\begin{eqnarray*}
&= & -\frac{4}{n^2} \frac{1}{\delta}\frac{\partial}{\partial \omega}
     Im(\frac{-2i sin(\omega(n-1)\delta/2) e^{i\omega(n-1)\delta/2} e^{i\omega \delta/2}}{-2i sin(\omega \delta/2) }\\
&=&  -\frac{4}{n^2} \frac{1}{\delta}\frac{\partial}{\partial \omega}
     \frac{sin(\omega (n-1)\delta/2}{sin\omega\delta/2} Im(e^{n i \omega \delta/2})
\end{eqnarray*}

This contribution to $<D_k^2>$ is 

$$ -\frac{4}{n^2} \frac{1}{\delta}\frac{\partial}{\partial \omega}
       \frac{sin(\omega (n-1)\delta/2)}{sin \omega\delta/2} sin \omega(n\delta/2))
       (1 - cos \omega \Delta)(A + \frac{B}{\omega^{\alpha}})$$

After evaluating the derivative, the imaginary part contribution is 
\begin{equation}
\frac{2}{n^2 sin^2(\omega \delta/2)} ( sin^2(n \omega \delta/2)- n sin(\omega(2n-1)\delta/2)sin(\omega \delta/2)) 
  (1 - cos \omega \Delta)(A + \frac{B}{\omega^{\alpha}})
\end{equation}

\section{The final formula}

Adding the real and imaginary contributions obtained in eq. (17) and (18)
\begin{eqnarray}
<D_k^2> &  = &   \int d\omega(1-cos\omega \Delta)(A + \frac{B}{\omega^{\alpha}}) [\frac{2}{n}  \\ 
        &  + &   \frac{4}{n}  \frac{cos(n \omega \delta/2) sin((n-1) \omega \delta/2)}{sin(\omega \delta/2)}
\nonumber\\
        &  + & \frac{2}{n^2 sin^2(\omega \delta/2)} ( sin^2(n \omega \delta/2)- n sin(\omega(2n-1)\delta/2)
                sin(\omega \delta/2))] \nonumber
\end{eqnarray}

\end{document}